\def\ieeecopyrightnoteversion{}
\documentclass{article}
\usepackage{spconf,amsmath,graphicx}

\usepackage{cite}
\usepackage{amsmath}
\usepackage{hyperref} %
\usepackage{tikz}
\usetikzlibrary{external}
\usetikzlibrary{calc}
\usepackage{microtype} %
\usepackage{pgfplots}
\pgfplotsset{compat=newest}
\usepackage{tabularx}
\usepackage{amssymb}
\usepackage{siunitx}
\usepackage{xcolor}
\usepackage{comment}
\usepackage{algorithm,algorithmic}
\usepackage{lipsum} 

\usepackage{etoolbox}
\usepackage{caption}

\ifdefined\ieeecopyrightnoteversion
\usepackage{lastpage}
\usepackage{eso-pic}
\usepackage{calc}
\newlength{\msize}
\newlength{\mpwidth}
\AddToShipoutPicture{
	\put(0,20){%
		\hspace{\msize}
		\begin{minipage}[c]{\mpwidth}
			\scriptsize
			\centering
			{\copyright 2021-2022 IEEE. Personal use of this material is permitted. Permission from IEEE must be obtained for all other uses, in any current or future media, including reprinting/republishing this material for advertising or promotional purposes, creating new collective works, for resale or redistribution to servers or lists, or reuse of any copyrighted component of this work in other works.}
		\end{minipage}%
	}
}
\fi

\ifdefined\websiteversion
\usepackage{lastpage}
\usepackage{eso-pic}
\usepackage{calc}
\newlength{\msize}

  \AddToShipoutPicture{
	\put(0,20){%
		\begin{minipage}{\paperwidth}
			\scriptsize
			\hspace{1in}\hspace{\oddsidemargin}
			IEEE International Conference on Acoustics, Speech and Signal Processing (ICASSP), May 2022, Singapore,
			\textsc{DOI}: \href{https://doi.org/10.1109/ICASSP43922.2022.9746559}{10.1109/ICASSP43922.2022.9746559}.
			\hfill \thepage\,/\,\pageref*{LastPage}
			\hspace{1in}\hspace{\oddsidemargin}
		\end{minipage}%
	}
}
\fi

\newcommand{\transpose}[1]{\ensuremath{#1^\mathrm{T}}}
\newcommand{\E}[1]{\mathbb{E} \left\{ #1 \right\}}
\newcommand{\sigIn}{x}
\newcommand{\sigInTotal}{\tilde{x}}
\newcommand{\sigOut}{d}
\newcommand{\sigOutNoisy}{y}
\newcommand{\sigOutNoisyTotal}{\tilde{y}}
\newcommand{\idxTime}{k}
\newcommand{\idxTimeAlt}{\kappa}
\newcommand{\ir}{h}
\newcommand{\irEst}{\hat{h}}
\newcommand{\vecIr}{\mathbf{h}}
\newcommand{\vecIrEst}{\mathbf{\hat{h}}}
\newcommand{\irLength}{\tilde{L}}
\newcommand{\irLengthEst}{L}
\newcommand{\sigNoise}{n}

\newcommand{\numSamplesTotal}{N_{\text{t}}}
\newcommand{\numSamplesFrame}{N_{\text{f}}}
\newcommand{\numSamplesLookahead}{N_{\text{a}}}
\newcommand{\numSamplesLookback}{N_{\text{b}}}

\newcommand{\numSources}{S}
\newcommand{\idxSource}{s}

\newcommand{\vecSigIn}{\mathbf{x}}
\newcommand{\vecProcessNoise}{\mathbf{q}}
\newcommand{\vecMeasurementNoise}{\mathbf{n}}

\newcommand{\varMeasurementNoise}{\sigma_n^2}

\newcommand{\sysDist}{D}
\newcommand{\sysDistMean}{\overline{D}}

\newcommand{\numIterations}{\mathcal{I}}

\newcommand{\mtxObservation}{\mathbf{C}}
\newcommand{\mtxStateCovariancePriori}{\mathbf{P}}
\newcommand{\mtxStateCovariancePosteriori}{\mathbf{V}}
\newcommand{\mtxStateCovarianceSmoothed}{\mathbf{\hat{V}}}
\newcommand{\mtxProcessNoise}{\boldsymbol{\Gamma}}
\newcommand{\mtxMeasurementNoise}{\boldsymbol{\Sigma}}
\newcommand{\mtxStateTransition}{\mathbf{A}}
\newcommand{\vecState}{\mathbf{z}}
\newcommand{\vecStateEstimate}{\boldsymbol{\mu}}
\newcommand{\vecStateEstimateSmoothed}{\boldsymbol{\hat{\mu}}}
\newcommand{\vecObservation}{\mathbf{y}}
\newcommand{\identity}{\mathbf{I}}

\newcommand{\mtxKalmanGain}{\mathbf{K}}
\newcommand{\mtxKalmanSmootherGain}{\mathbf{J}}

\newcommand{\numObservations}{N}

\newcommand{\paramSet}{\boldsymbol{\theta}}

\definecolor{mycolor2}{rgb}{0.80000,0.02745,0.11765}%

\title{Towards Faster Continuous Multi-Channel HRTF Measurements Based on Learning System Models}
\name{Tobias Kabzinski and Peter Jax}
\address{Institute of Communication Systems (IKS), RWTH Aachen University, Aachen, Germany\\
	\textit{\{kabzinski,jax\}@iks.rwth-aachen.de}}

\makeatletter
\def\section{%
	\vspace{-0.5ex} %
	\@startsection {section}
	{1}
	{\z@ }
	{-3.5ex \@plus -1ex \@minus -.2ex}
	{1.5ex \@plus .2ex} %
	{\normalfont \Large \bfseries }%
}
\makeatother

\begin{document}
\ninept

\makeatletter
\hypersetup{pdftitle={\@title},pdfauthor={Tobias Kabzinski, Peter Jax}, hidelinks}
\makeatother

\addtolength{\abovecaptionskip}{-2pt} %
\addtolength{\belowcaptionskip}{-6pt}
\pretocmd{\caption}{\vspace*{-1.5ex}}{}{} %

\newlength{\mydisplayskip}
\setlength{\mydisplayskip}{5pt minus 2pt}
\setlength{\abovedisplayskip}{\mydisplayskip}
\setlength{\belowdisplayskip}{\mydisplayskip}
\setlength{\abovedisplayshortskip}{0pt plus 2pt minus 2pt}
\setlength{\belowdisplayshortskip}{\mydisplayskip}

\maketitle
\begin{abstract}
Measuring personal head-related transfer functions (HRTFs) is essential in binaural audio. Personal HRTFs are not only required for binaural rendering and for loudspeaker-based binaural reproduction using crosstalk cancellation, but they also serve as a basis for data-driven HRTF individualization techniques and psychoacoustic experiments. Although many attempts have been made to expedite HRTF measurements, the rotational velocities in today's measurement systems remain lower than those in natural head movements. To cope with faster rotations, we present a novel continuous HRTF measurement method. This method estimates the HRTFs offline using a Kalman smoother and learns state-space parameters, including the system model, on short signal segments, utilizing the expectation maximization algorithm. We evaluated our method in simulated single-channel and multi-channel measurements using a rigid sphere HRTF model. Comparing with conventional methods, we found that the system distances are improved by up to $\SI{30}{dB}$. %

\end{abstract}

\begin{keywords}
expectation maximization, head-related transfer functions, spatial audio, system identification
\end{keywords}
\section{Introduction}

Head-related transfer functions (HRTFs) have several applications in spatial audio. An HRTF describes the (normalized) transfer function from a sound source at a given direction to the listener's ear \cite{moller1992fundamentals}. HRTFs are essential to render virtual sound sources and virtual acoustic scenes for headphone reproduction \cite{begault1994sound,vorlander2007auralization}. In addition, loudspeaker-based binaural reproduction with a crosstalk cancellation system requires HRTFs to compute the crosstalk cancellation filters \cite{lentz2006dynamic,masiero2014framework}. It was found that using non-individualized HRTFs can degrade the localization performance in both headphone-based \cite{wenzel1993localization} and loudspeaker-based \cite{majdak2013sound} reproduction, which highlights the importance of HRTF individualization techniques (cf. \cite{guezenoc2018hrtf,sunder2015natural} for overviews). These techniques aim at providing individualized HRTFs, for example, based on anthropometric data. Although the interest in these HRTF individualization techniques grows, acoustic measurements represent the standard way to obtain reference data for individualization techniques and for psychoacoustic experiments. %

Measuring HRTFs quickly and precisely has been studied for many years (cf. \cite{li2020measurement,enzner2013trends} for reviews). Two main approaches can be distinguished: stop-and-go measurements and continuous measurements. In the former approach, a transfer function is identified separately for each direction. In this case, exponential-sweep excitation and deconvolution are commonly used. This allows to cope with weak loudspeaker nonlinearities \cite{muller2001transfer,farina2007advancements}. To reduce the measurement duration it has been proposed to interleave the sweeps from multiple loudspeakers \cite{majdak2007multiple}. In the latter approach, the subject rotates continuously during the measurement. Here, noise signals or perfect sequences (PSEQs) serve as excitation signals while adaptive filters continuously estimate the transfer function(s) from one or more loudspeakers to the ear \cite{enzner2008analysis,enzner20093d,antweiler2009perfect}. Typically, the normalized least-mean-square (NLMS) algorithm is employed to obtain time-domain representations of the HRTFs, i.e., head-related impulse responses (HRIRs). Recently, exponential-sweep measurements with slow continuous subject rotations were investigated \cite{richter2019influence}. Moreover, it was shown that deconvolution and NLMS-based system identification are algorithmically equivalent for a single-loudspeaker setup\cite{kuhl2018joint}.

Over the past few years, the interest in continuous HRTF measurements has grown. For instance, they were used to measure the HRTF database in \cite{brinkmann2019cross}. Besides, continuous measurements are attractive for measurements with unconstrained subject movement, such as in \cite{he2018fast,nagel2018acoustic}. While the NLMS algorithm and its variants in \cite{he2018fast} are easy to implement, more powerful adaptive filtering algorithms exist. Specifically, one drawback of the NLMS algorithm is that it lacks a model which describes how the impulse response evolves over time. This might be a reason why the rotational velocities at which a precise measurement result can be achieved remain low. Ideally, a measurement system would be able to identify HRTFs at natural head rotation velocities, which can reach almost \SI[number-unit-product={}]{600}{\degree \per \second} \cite{zangemeister1981dynamics}.

The HRTF measurement problem is a system identification problem for time-variant systems, which is similarly encountered in acoustic echo control (AEC). In AEC, it has been suggested to employ a Kalman filter to estimate the impulse response or transfer function, which is, in the state-space approach, described by the state vector \cite{enzner2006frequency,enzner2010bayesian}. In these works, the system model comprises a first-order Markov model with a scalar fading factor to model the time-variance. Performance bounds resulting from this assumption were studied in \cite{fabry2020steady} for the time-domain Kalman filter. In \cite{malik2010online}, online estimates for the process noise and measurement noise covariance matrix of the diagonalized frequency-domain Kalman filter were derived assuming a first-order Markov model with scalar fading factor.

The main idea of this paper is to learn a state-space system model to accurately describe the evolution of the HRIRs over time during a continuous rotation. Therefore, the scalar fading factor is replaced by a generic state transition matrix. We suggest to learn the system model parameters, i.e., the generic state transition matrix and the process noise covariance matrix, using the expectation maximization (EM) algorithm for linear dynamical systems (LDSs) \cite{ghahramani1996parameter,bishop2006pattern}. The measurement noise covariance can also be learned this way. Using the LDS-EM algorithm has two consequences. Firstly, the more general state-transition model increases the computational burden compared to the model with a scalar fading factor. Secondly, the LDS-EM algorithm requires offline processing as future samples must be accessible. While these two restrictions prohibit low-latency real-time processing, as required e.g., in AEC, they do not pose a restriction for continuous HRTF measurements. Here, the processing can be conducted in an offline manner after recording the 
signals.

\section{Signal Model}
\label{sec:signalmodel}

To formalize the HRTF measurement problem, we consider, without loss of generality, a multiple-input single-output (MISO) system with a single microphone at the left ear. Let $\sigIn_{\idxSource}\!\left( \idxTime \right)$ denote the length-$\numObservations$ excitation signal at loudspeaker $\idxSource$ and time sample $\idxTime$. Furthermore, $\ir_{\idxTime,\idxSource}\!\left( \idxTimeAlt \right)$ denotes the $\idxTimeAlt$-th sample of the time-variant impulse response at sample $\idxTime$ between the $\idxSource$-th source and the microphone. Then, the clean output signal is given by
\begin{equation}
	\label{eqn:outputsignal}
	\sigOut \left( \idxTime \right) 
	= \sum_{\idxSource=1}^{\numSources} \sum_{\idxTimeAlt=0}^{\irLength-1} 
	\sigIn_{\idxSource}\!\left( \idxTime - \idxTimeAlt \right) \ir_{\idxTime,\idxSource}\!\left( \idxTimeAlt \right).
\end{equation}
Here, $\numSources$ is the number of loudspeakers and $\irLength$ denotes the impulse response length. Moreover, it is assumed that the observable microphone signal $\sigOutNoisy\!\left( \idxTime \right)$ is noisy, i.e., it is superimposed by white Gaussian noise to model the microphone self-noise: $\sigOutNoisy\!\left( \idxTime \right) = \sigOut\!\left( \idxTime \right) + \sigNoise\!\left( \idxTime \right)$.

The goal of continuous HRTF measurements is to obtain HRTF or HRIR estimates $\irEst_{\idxTime,\idxSource}\!\left( \idxTimeAlt \right)$ at each time $\idxTime$ from each loudspeaker, which are potentially placed at different elevation angles, as in \cite{enzner20093d}. Then, the HRIRs can be mapped to the relative loudspeaker-subject orientations, recorded with a head tracker in parallel to the measurement. To facilitate the state-space description, we define a compact notation. Let 
$
	\vecSigIn_{\idxTime,\idxSource}
	= \transpose{
		\left[
		\sigIn_{\idxSource}\!\left( \idxTime \right), \ldots, \sigIn_{\idxSource}\!\left( \idxTime - \irLengthEst + 1\right)
		\right]
	}
$
be the vector including the $\irLengthEst$ most recent input signal samples of the $\idxSource$-th loudspeaker signal at time $\idxTime$. Then, all input signals can be stacked into one vector 
$
	\vecSigIn_{\idxTime}
	= \transpose{ \left[
		\transpose{ \vecSigIn_{\idxTime,1}}, \ldots, \transpose{ \vecSigIn_{\idxTime,\numSources}}
		\right]
	}
$.
Similarly, we define the time-dependent $\irLengthEst$-tap impulse response vector from loudspeaker $\idxSource$ to the microphone as
$
	\vecIr_{\idxTime, \idxSource}
	= \transpose{ 
		\left[
		\ir_{\idxTime,\idxSource}\!\left( 0 \right), \ldots, \ir_{\idxTime,\idxSource}\!\left( \irLengthEst-1 \right)
		\right]
	}$.
We want to estimate the state vector $\vecState_{\idxTime}$, which comprises all the $\numSources \irLengthEst$ impulse responses coefficients, i.e., 
$
	\vecState_{\idxTime}
	= \transpose{ \left[ 
		\transpose{ \vecIr_{\idxTime, 1} }, \ldots, \transpose{ \vecIr_{\idxTime, \numSources} }
		\right]}$.

In contrast to \cite{enzner2010bayesian} (and \cite{lopes2015random} for the multi-channel case), we propose to replace the scalar fading factor by a generic state transition matrix $\mtxStateTransition$ and obtain the state equation
\begin{equation}
	\label{eqn:stateequation}
	\vecState_{\idxTime} = \mtxStateTransition \vecState_{\idxTime-1} + \vecProcessNoise_{\idxTime}.
\end{equation}
The process noise $\vecProcessNoise_{\idxTime}$ is assumed to be distributed by a zero-mean multivariate normal distribution with covariance matrix $\mtxProcessNoise$. The observation equation of the state-space representation is given by
\begin{equation}
	\label{eqn:observationequation}	
	\vecObservation_{\idxTime} = \sigOutNoisy\!\left( \idxTime \right) 
	= \mtxObservation_{\idxTime} \vecState_{\idxTime} + \vecMeasurementNoise_{\idxTime}
	= \transpose{ \vecSigIn_{\idxTime}} \vecState_{\idxTime} + \sigNoise\!\left( \idxTime  \right),
\end{equation}
where the $1 \times \irLengthEst\numSources$ observation matrix $\mtxObservation_{\idxTime} = \transpose{ \vecSigIn_{\idxTime}}$ implements the convolution operation and summation over all the loudspeakers in~\eqref{eqn:outputsignal}. The measurement noise $\vecMeasurementNoise_{\idxTime}$ is assumed to be normally distributed with zero mean and variance $\mtxMeasurementNoise = \varMeasurementNoise$.

\section{Proposed Measurement Method}
\label{sec:proposedsystem}
Before proceeding with the proposed processing scheme in Section~\ref{ssec:processing}, we first present the required components, namely, the Kalman smoother and the EM algorithm for learning parameters in state-space systems.

\subsection{Kalman Filter and Kalman Smoother}
\label{ssec:kalman}

Given the state equation \eqref{eqn:stateequation} and the observation equation \eqref{eqn:observationequation}, the Kalman \emph{filter} constitutes the best linear minimum-mean-squared-error (MMSE) estimator of the state $\vecState_{\idxTime}$ at any time step $\idxTime$ when only observations up to and including this time step $\idxTime$ are accessible \cite{grewal2014kalman,bishop2006pattern}. The following update equations for the \textit{a priori} state covariance matrix $\mtxStateCovariancePriori_{\idxTime}$, the Kalman gain $\mtxKalmanGain_{\idxTime}$, the (forward) state estimate $\vecStateEstimate_{\idxTime}$, and the \textit{a posteriori} state covariance matrix $\mtxStateCovariancePosteriori_{\idxTime}$ need to be conducted recursively for $\idxTime=1, \ldots, \numObservations$ to obtain HRIR estimates in $\vecStateEstimate_{\idxTime}$ at each time step:
\begin{subequations}
	\begin{align}
	\begin{split}
	\label{eqn:kalmanfilterpriori}
	\mtxStateCovariancePriori_{\idxTime-1} 
	&= \begin{cases}
		\mtxStateCovariancePriori_{0} & \text{if } \idxTime = 1, \\
		\mtxStateTransition \mtxStateCovariancePosteriori_{\idxTime-1} \transpose{\mtxStateTransition} + \mtxProcessNoise & \text{else},
	\end{cases}
	\end{split} \\
	\begin{split}
	\label{eqn:kalmanfiltergain}	
	\mtxKalmanGain_{\idxTime}
	&= \mtxStateCovariancePriori_{\idxTime-1} \transpose{\mtxObservation_{\idxTime}}
	\left( 
	\mtxObservation_{\idxTime} \mtxStateCovariancePriori_{\idxTime-1} \transpose{\mtxObservation_{\idxTime}} + \mtxMeasurementNoise
	\right)^{-1},
	\end{split} \\
	\begin{split}
	\label{eqn:kalmanfilterstate}	
	\vecStateEstimate_{\idxTime}
	&= \begin{cases}
		\vecStateEstimate_{0} & \text{if } \idxTime = 1, \\
		\mtxStateTransition \vecStateEstimate_{\idxTime-1}
		+ \mtxKalmanGain_{\idxTime} \left( \vecObservation_{\idxTime} - \mtxObservation_{\idxTime} \mtxStateTransition \vecStateEstimate_{\idxTime-1} \right) & \text{else},
	\end{cases}	
	\end{split} \\
	\begin{split}
	\label{eqn:kalmanfilterposteriori}	
	\mtxStateCovariancePosteriori_{\idxTime}
	&= \left( \identity - \mtxKalmanGain_{\idxTime} \mtxObservation_{\idxTime} \right) \mtxStateCovariancePriori_{\idxTime-1}.
	\end{split}
	\end{align}
\end{subequations}
Here, $\identity$ is the identity matrix. An initial mean state vector $\vecStateEstimate_{0}$ and an associated initial \textit{a priori} covariance matrix $\mtxStateCovariancePriori_{0}$ are required.

After Kalman filtering, Kalman \emph{smoothing} is conducted to obtain improved (backward) state estimates $\vecStateEstimateSmoothed_{\idxTime}$ and improved (smoothed) state covariance matrices $\mtxStateCovarianceSmoothed_{\idxTime}$ since future samples are accessible in offline processing. So, the recursion, including the Kalman smoother gain $\mtxKalmanSmootherGain_{\idxTime}$, with the below update equations \cite{bishop2006pattern}, starts from the end of the sequence, i.e., from $\idxTime = \numObservations,\ldots,1$:
\begin{subequations}
	\begin{align}
		\begin{split}
	\label{eqn:kalmansmoothergain}	
	\mtxKalmanSmootherGain_{\idxTime}
	&= \mtxStateCovariancePosteriori_{\idxTime} \transpose{ \mtxStateTransition } \mtxStateCovariancePriori_{\idxTime}^{-1},
\end{split} \\
\begin{split}
	\label{eqn:kalmansmootherstate}	
	\vecStateEstimateSmoothed_{\idxTime} 
	&= \begin{cases}
		\vecStateEstimate_{\idxTime} & \text{if } \idxTime = \numObservations, \\
		\vecStateEstimate_{\idxTime} + \mtxKalmanSmootherGain_{\idxTime} 
		\left( \vecStateEstimateSmoothed_{\idxTime+1} - \mtxStateTransition \vecStateEstimate_{\idxTime} \right) & \text{else},
	\end{cases}
\end{split} \\
\begin{split}
	\label{eqn:kalmansmootherposteriori}
	\mtxStateCovarianceSmoothed_{\idxTime} 
	&= \begin{cases}
		\mtxStateCovariancePosteriori_{\idxTime} & \text{if } \idxTime = \numObservations, \\
		\mtxStateCovariancePosteriori_{\idxTime} + \mtxKalmanSmootherGain_{\idxTime}
		\left( \mtxStateCovarianceSmoothed_{\idxTime+1} - \mtxStateCovariancePriori_{\idxTime} \right) \transpose{ \mtxKalmanSmootherGain_{\idxTime}}
		& \text{else}.
	\end{cases}
	\end{split}
\end{align}
\end{subequations}
Finally, the HRIR estimates $\vecIrEst_{\idxTime, \idxSource}$ are extracted from $\vecStateEstimateSmoothed_{\idxTime}$ for all $\idxTime$.

\subsection{Parameter~Learning~Using~Expectation~Maximization~(EM)}
\label{ssec:em}

Implementing the Kalman filter and Kalman smoother necessitates assumptions about $\mtxStateTransition, \mtxProcessNoise$ and $\mtxMeasurementNoise$. Traditionally, the process noise covariance matrix $\mtxProcessNoise$ is approximated as a diagonal matrix, and it is computed via recursive averaging, e.g., in \cite{paleologu2013study} and \cite{liebich2017time}. As a result, $\mtxProcessNoise$ becomes a time-dependent matrix $\mtxProcessNoise_{\idxTime}$. Similarly, the scalar fading factor in the Markov model is either simply set to one or set heuristically, e.g., in \cite{paleologu2013study} and \cite{enzner2006frequency,enzner2010bayesian}, respectively. The measurement noise $\varMeasurementNoise$ could be estimated online \cite{strutz2019estimation} and then also becomes time dependent. In contrast to these approaches, we suggest to learn all the unknown parameters $\mtxStateTransition, \mtxProcessNoise, \mtxMeasurementNoise$, and the initial values $\vecStateEstimate_{0}, \mtxStateCovariancePriori_{0}$ utilizing the LDS-EM algorithm \cite{ghahramani1996parameter,bishop2006pattern}.

Given sequences of the input signals $\sigIn_{\idxSource}\!\left( \idxTime \right), \idxSource=1,\ldots,\numSources$ and the microphone signal $\sigOutNoisy\! \left( \idxTime \right)$, both for $\idxTime=1,\ldots,\numObservations$, the EM algorithm provides parameter estimates by maximizing the log-likelihood of these parameters for the given input sequence $\vecSigIn_{\idxTime}$ and the observation sequence $\vecObservation_{\idxTime}$. One iteration of the EM algorithm consists of the following two steps. In the E-step, the Kalman smoother estimates are evaluated for a given set of parameters $\paramSet = \left\{ \mtxStateTransition, \mtxProcessNoise, \mtxMeasurementNoise, \vecStateEstimate_{0}, \mtxStateCovariancePriori_{0} \right\}$. Whereas in the M-step, the parameters in the set are updated using the quantities highlighted by the $^\star$ symbol below, i.e.,
\begin{equation}
	\label{eqn:mstepstatetransition}
	\mtxStateTransition^{\star}
	= \left( \sum_{\idxTime=2}^{\numObservations}	\E {	\vecState_\idxTime \transpose{\vecState_{\idxTime-1}} } \right)
	\left( \sum_{\idxTime=2}^{\numObservations} \E{  \vecState_{\idxTime-1} \transpose{\vecState_{\idxTime-1}} } \right)^{-1},
\end{equation}
\begin{align}
	\label{eqn:mstepprocessnoise}
	\mtxProcessNoise^{\star}
	= \frac{1}{\numObservations-1} \sum_{\idxTime=2}^{\numObservations} 
	\E { \vecState_\idxTime \transpose{ \vecState_\idxTime} }
	- \E {  \vecState_\idxTime \transpose{\vecState_{\idxTime-1}}   }\transpose{{\mtxStateTransition^{\star}}} \nonumber \\
	- \mtxStateTransition^{\star} \E {  \vecState_{\idxTime-1} \transpose{\vecState_\idxTime} } 
	+ \mtxStateTransition^{\star} \E { \vecState_{\idxTime-1} \transpose{\vecState_{\idxTime-1}}  } \transpose{{\mtxStateTransition^{\star}}}, 		
\end{align}%
\begin{align}
	\label{eqn:mstepmeasurementnoise}
	\mtxMeasurementNoise^{\star}\!
	&=\!\frac{1}{\numObservations} \sum_{\idxTime=1}^{\numObservations}
	\left[ \sigOutNoisy \left( \idxTime \right) \right]^2
	\!-\!2 \mtxObservation_{\idxTime} \vecStateEstimateSmoothed_{\idxTime} \sigOutNoisy\!\left( \idxTime \right)
	\!+\! \mtxObservation_{\idxTime} \E{\vecState_{\idxTime} \transpose{\vecState_{\idxTime}}} \transpose{ \mtxObservation_{\idxTime} },
\end{align}%
\vspace{-1.5em}
\begin{tabularx}{\columnwidth}{@{}XX@{}}
\begin{equation}
	\label{eqn:mstepinitstate}
	\vecStateEstimate_{0}^{\star} = \vecStateEstimateSmoothed_1,
\end{equation} & 
\begin{equation}
	\label{eqn:mstepinitcov}	
	\mtxStateCovariancePriori_{0}^{\star} = \mtxStateCovarianceSmoothed_1.
\end{equation}
\end{tabularx}
To evaluate \eqref{eqn:mstepstatetransition} to \eqref{eqn:mstepmeasurementnoise}, the equations $\E{\vecState_{\idxTime} \transpose{\vecState_{\idxTime}}} = \mtxStateCovarianceSmoothed_{\idxTime}	+ \vecStateEstimateSmoothed_{\idxTime} \transpose{\vecStateEstimateSmoothed_{\idxTime}}$ and $\E{\vecState_{\idxTime} \transpose{\vecState_{\idxTime-1}}} = \mtxStateCovarianceSmoothed_{\idxTime} \transpose{\mtxKalmanSmootherGain_{\idxTime-1}} + \vecStateEstimateSmoothed_{\idxTime} \transpose{\vecStateEstimateSmoothed_{\idxTime-1}}$
are used \cite{bishop2006pattern}. After E-step and M-step, the E-step of the next iteration is done with the updated parameters, and then another M-step can be conducted. This process is repeated $\numIterations$ times. Note that, in contrast to the solution in \cite{bishop2006pattern}, the observation matrix is time-dependent and deterministic.

\subsection{Processing Scheme}
\label{ssec:processing}

Let $\sigInTotal_{\idxSource}\!\left( \idxTime \right), \idxSource=1,\ldots, \numSources$ and $\sigOutNoisyTotal\!\left( \idxTime \right)$, both for $\idxTime = 0, \ldots, \numSamplesTotal-1$, denote the length-$\numSamplesTotal$ input signals and the microphone signal recorded during a continuous HRTF measurement, respectively. In theory, we could conduct the Kalman smoothing and parameter learning steps described above using these signals in order to obtain \emph{one set} of parameters for the entire sequence. In practice, however, the memory required to save the quantities needed becomes easily unmanageable. For instance, a setup with $\numSources = 3$ loudspeakers, $\irLengthEst=192$ estimated HRIR coefficients per loudspeaker, and $\numSamplesTotal = 480000$ samples for a 20-second recording at a sampling rate of \SI{24}{kHz} would require at least \SI{1.2}{TB} of memory. This size is dominated by the two matrices $\mtxStateCovarianceSmoothed_{\idxTime}, \mtxKalmanSmootherGain_{\idxTime}$ of size $\irLengthEst \numSources \times \irLengthEst \numSources$ saved at each of the $\numSamplesTotal$ time steps.

To resolve the memory requirement issue, we suggest to learn independently \emph{separate sets} of parameters on shorter signal excerpts, denoted as segments in the following. Each segment shall consist of $\numSamplesFrame$ samples of a current frame, $\numSamplesLookback$ lookback samples before the frame, and $\numSamplesLookahead$ lookahead samples after the frame. This results in $\numObservations = \numSamplesFrame + \numSamplesLookback + \numSamplesLookahead$ samples per segment. For each segment, $\numIterations$ iterations of the LDS-EM algorithm are conducted. With the latest parameter set obtained, the Kalman smoothing is done once more to obtain HRIR estimates $\vecIrEst_{\idxTime, \idxSource}$. Only the estimates obtained within the $\numSamplesFrame$ samples of the frame are regarded as the final estimates. We exclude the, potentially inaccurate, estimates obtained during the convergence phase in the lookback samples, and we exclude the estimates in the lookahead samples, where the Kalman smoother does not yield much better results than the Kalman filter\footnote{Exceptions from this procedure are required in the first and in the last segment such that HRIR estimates are obtained at every time sample $\idxTime$. In the first and in the last frame, the HRIR estimates do not come from the frame alone but also from the lookback and/or lookahead part(s).}. Having processed one segment, the next segment is obtained by sliding the length-$\numObservations$ excerpt (window) $\numSamplesFrame$ samples into the future.

\section{Evaluation and Results}
\label{sec:results}
We carried out various experiments to analyze the system model parameters learned by the proposed method and to compare against state-of-the-art continuous HRTF measurement methods.
\subsection{Setup and Metrics}

To assess the performance of a system identification algorithm for time-variant systems, it seems infeasible to measure thousands of HRTFs corresponding to different source positions. Moreover, a truly time-invariant measurement environment is difficult to realize. Thus, we decided to simulate the measurements as follows. To avoid any kind of HRTF interpolation, which might introduce additional errors, we consider the "HRTF" of a rigid sphere. Its transfer functions are structurally similar to human HRTFs with its deep notches, but the rigid sphere transfer functions can be calculated analytically for any angle of incidence \cite{duda1998range}. Therefore, they provide a ground truth to compute a metric at each sample to evaluate the system identification algorithm---not the imperfections of hardly reproducible real acoustic measurements. The time-variant HRIRs are modeled according to \cite{urbanietz2018binaural} for head rotations of different but constant velocities. We consider either one loudspeaker in the horizontal plane or three loudspeakers at elevations \SI[number-unit-product={}]{0}{\degree}, \SI[number-unit-product={}]{15}{\degree}, and \SI[number-unit-product={}]{30}{\degree} in an anechoic environment. The loudspeaker-sphere distance is \SI{1.5}{m}, and the sampling rate is set to $\SI{24}{kHz}$.  White Gaussian noise is added to the microphone signal to achieve an SNR of \SI{60}{dB} to simulate a typical measurement noise setting \cite{muller2001transfer}. The estimated HRIRs are \SI{8}{ms} long, i.e., $\irLengthEst = 192$. %

Comparisons of the following combinations of excitation signal and identification algorithm were conducted. Firstly, the NLMS algorithm with stepsize one was considered for the HRIR estimation and white noise (WN) was used as the excitation signal, similarly to \cite{enzner20093d}. Secondly, the white noise was replaced by a periodically repeated perfect sequence. The period length was set to $\irLengthEst \numSources$. This way, initial convergences to the true impulse response within $2 \irLengthEst \numSources$ samples is guaranteed for a time-invariant system \cite{antweiler2008multi-channel}, and hence a good tracking performance for a time-variant system is expected, similarly to \cite{antweiler2009perfect}. Thirdly, the NLMS algorithm was replaced by a Kalman filter (KF) with $\mtxStateTransition = \identity$, as in \cite{paleologu2013study}. The process noise was estimated by exponential smoothing with a time constant of \SI{50}{ms}, similarly to \cite{liebich2017time}. The measurement noise variance was set to $\mtxMeasurementNoise = \varMeasurementNoise = 10^{-6}$ to match the true SNR. Lastly, the proposed method was considered in combination with PSEQ excitation. For initialization, we chose $\mtxStateTransition = \identity$, $\mtxProcessNoise = 10^{-7} \cdot \identity$, $\varMeasurementNoise = 0.01$, $\vecStateEstimate_{0} = \mathbf{0}$, and $\mtxStateCovariancePriori_{0} = \identity$. We set frame length, lookback and lookahead to $\numSamplesFrame\!=\!\numSamplesLookback\!=\!\numSamplesLookahead\!=\!1200$ (\SI{50}{ms}), resulting in \SI{150}{ms} segments, a compromise for the rotational velocities considered here.

To quantify the quality of the HRIR estimates, the relative system distance, or normalized misalignment, is computed per loudspeaker, i.e.,
$
	\sysDist_{\idxSource}\!\left( \idxTime \right)
	= 10 \log_{10} \left( 
	\| \vecIr_{\idxTime, \idxSource} - \vecIrEst_{\idxTime, \idxSource} \|^2
	/ \|  \vecIr_{\idxTime, \idxSource} \|^2 
	\right)
$. Here, the estimated HRIRs are zeropadded to match the length $\irLength = 315$ of the reference HRIRs. Further, we define
$
	\sysDistMean = \frac{1}{\numSamplesTotal - 2 \numSources \irLengthEst} \sum_{\idxTime=2 \numSources \irLengthEst}^{\numSamplesTotal-1} \sum_{\idxSource=1}^{\numSources} \sysDist_{\idxSource}\!\left( \idxTime \right)
$	
as the average system distance of all channels excluding the $2 \numSources \irLengthEst$ samples in the initial convergence phase \cite{antweiler2008multi-channel}. The length $\numSamplesTotal$ of each simulation was chosen to cover, due to symmetry, a \SI[number-unit-product={}]{180}{\degree} rotation after the initial convergence phase.

\subsection{Experiments and Results}

\begin{figure}%
	\centering
	\input{images/detailAnalysisJoint.tikz}
	\caption{Absolute values of $\mtxStateTransition$ and $\mtxProcessNoise$ in the first segment of a \SI[number-unit-product={}]{360}{\degree \per \second} rotation. The color scaled is warped using $f(x) = \sqrt[10]{\left|x\right|}$.}
	\vspace{-1em}
	\label{fig:detailAnalysisJoint}
\end{figure}

Fig.~\ref{fig:detailAnalysisJoint} visualizes the two learned parameters $\mtxStateTransition$ and $\mtxProcessNoise$ after $\numIterations=10$ iterations of the LDS-EM algorithm for a single-channel experiment. Note that warped color scales are used to improve the visibility of small values. In the state transition matrix, the diagonal values are large and many off-diagonal values are nonzero. This implies that one HRIR coefficient at the next time step is well modeled as a weighted version of the same coefficient at the previous time step---with a weight close to one on the main diagonal---and a linear combination of other HRIR coefficients with smaller weights. These weights are larger around the direct-path samples near the base delay of \SI{4.4}{ms}, corresponding to the loudspeaker-microphone distance of \SI{1.5}{m}. Conversely, the direct-path samples do not contribute to the weights of the samples before \SI{4}{ms} and after \SI{5}{ms}. In $\mtxProcessNoise$ nonzero cross-covariances appear, indicating that the impulse response coefficients do not change independently of each other. Structurally, both matrices look diagonally dominant---similar to their initial values. This seems plausible since the EM algorithm is likely to have converged, after 10 iterations, to somewhere close to a local maximum not too far from the initialization.

\begin{figure}%
	\centering
	\definecolor{mycolor1}{rgb}{0.96471,0.65882,0.00000}%
\definecolor{mycolor2}{rgb}{0.80000,0.02745,0.11765}%
\definecolor{mycolor3}{rgb}{0.34118,0.67059,0.15294}%
\definecolor{mycolor4}{rgb}{0.00000,0.32941,0.62353}%
\begin{tikzpicture}

\begin{axis}[%
width=0.808\columnwidth,
height=7.5cm,
at={(0\columnwidth,0cm)},
scale only axis,
unbounded coords=jump,
xmin=0.000000,
xmax=1.015958,
xtick={  0, 0.1, 0.2, 0.3, 0.4, 0.5, 0.6, 0.7, 0.8, 0.9,   1},
xlabel style={font=\color{white!15!black}},
xlabel={time / $\si{\second}$},
ymin=-70.000000,
ymax=10.000000,
ylabel style={font=\color{white!15!black}},
ylabel={system distance $\sysDist_1\!\left( \idxTime \right)$ / \si{dB}},
axis background/.style={fill=white},
xmajorgrids,
ymajorgrids,
legend style={at={(0.03,0.97)}, anchor=north west, legend cell align=left, align=left, draw=white!15!black},
footnotesize,legend style={font=\tiny,at={(axis cs: 0.08, 10)}},width=0.85\columnwidth, line join=round,width=7.0cm,height=3.2cm,legend columns=2,transpose legend,label style={font=\scriptsize},tick label style={font=\scriptsize}
]
\addplot [color=white!60!black, dotted, line width=1.0pt]
  table[]{images/data/sysDistOverTimeSingleChannel-1.tsv};
\addlegendentry{NLMS (WN)}

\addplot [color=mycolor1, dashdotted, line width=1.0pt]
  table[]{images/data/sysDistOverTimeSingleChannel-2.tsv};
\addlegendentry{NLMS}

\addplot [color=mycolor2, dashdotted, line width=1.0pt]
  table[]{images/data/sysDistOverTimeSingleChannel-3.tsv};
\addlegendentry{NLMS (shifted)}

\addplot [color=mycolor3, dashed, line width=1.0pt]
  table[]{images/data/sysDistOverTimeSingleChannel-4.tsv};
\addlegendentry{KF}

\addplot [color=mycolor4, line width=1.0pt]
  table[]{images/data/sysDistOverTimeSingleChannel-5.tsv};
\addlegendentry{Proposed: $\numIterations=10$}

\addplot [color=black, dotted, line width=1.0pt]
  table[]{images/data/sysDistOverTimeSingleChannel-6.tsv};
\addlegendentry{TVI}

\end{axis}
\end{tikzpicture}%
	\caption{Single-channel setup with rotational velocity \SI[number-unit-product={}]{180}{\degree \per \second}.}
	\vspace{-1em}
	\label{fig:sysDistOverTimeSingleChannel}
\end{figure}

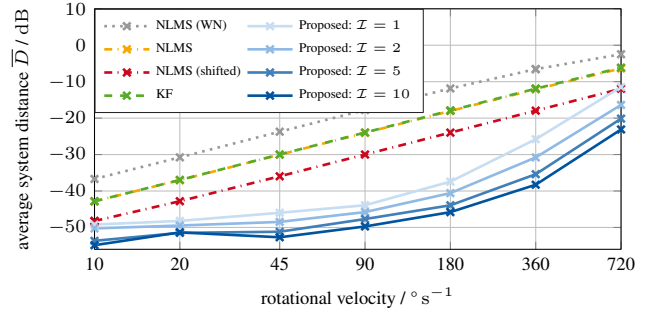
\begin{figure}%
	\centering
	\definecolor{mycolor1}{rgb}{0.96471,0.65882,0.00000}%
\definecolor{mycolor2}{rgb}{0.80000,0.02745,0.11765}%
\definecolor{mycolor3}{rgb}{0.34118,0.67059,0.15294}%
\definecolor{mycolor4}{rgb}{0.78039,0.86667,0.94902}%
\definecolor{mycolor5}{rgb}{0.55686,0.72941,0.89804}%
\definecolor{mycolor6}{rgb}{0.25098,0.49804,0.71765}%
\definecolor{mycolor7}{rgb}{0.00000,0.32941,0.62353}%
\begin{tikzpicture}

\begin{axis}[%
width=0.808\columnwidth,
height=7.5cm,
at={(0\columnwidth,0cm)},
scale only axis,
xmode=log,
xmin=10,
xmax=720,
xtick={10,20,45,90,180,360,720},
xticklabels={{10},{20},{45},{90},{180},{360},{720}},
xminorticks=true,
xlabel style={font=\color{white!15!black}},
xlabel={rotational velocity / $\si{\degree\per\second}$},
ymin=-56,
ymax=10,
ylabel style={font=\color{white!15!black}},
ylabel={average system distance $\sysDistMean$ / \si{dB}},
axis background/.style={fill=white},
xmajorgrids,
xminorgrids,
ymajorgrids,
legend style={at={(0.03,0.97)}, anchor=north west, legend cell align=left, align=left, draw=white!15!black},
footnotesize,legend style={font=\tiny,at={(axis cs: 10, 10)}},width=0.85\columnwidth, line join=round,width=7.0cm,height=3.2cm,legend columns = 4, transpose legend,label style={font=\scriptsize},tick label style={font=\scriptsize},
]
\addplot [color=white!60!black, dotted, line width=1.0pt, mark=x, mark options={solid, white!60!black}]
  table[row sep=crcr]{%
10	-36.693868120183\\
20	-30.7683118084579\\
45	-23.7251882156325\\
90	-17.7673033727997\\
180	-11.8570506171458\\
360	-6.56407511029057\\
360	-6.56407511029057\\
720	-2.47002422845491\\
};
\addlegendentry{NLMS (WN)}

\addplot [color=mycolor1, dashdotted, line width=1.0pt, mark=x, mark options={solid, mycolor1}]
  table[row sep=crcr]{%
10	-42.8383870037783\\
10	-42.838387003779\\
20	-37.0112675837699\\
45	-30.0381074317587\\
90	-24.0408239825259\\
180	-18.0433564131947\\
360	-12.0945868569731\\
720	-6.36933534814272\\
};
\addlegendentry{NLMS}

\addplot [color=mycolor2, dashdotted, line width=1.0pt, mark=x, mark options={solid, mycolor2}]
  table[row sep=crcr]{%
10	-48.2361169035484\\
10	-48.2361169035481\\
20	-42.8007703677876\\
45	-35.9643990469282\\
90	-29.9941531906288\\
180	-23.9764703957282\\
360	-17.9516137993033\\
720	-11.9361705079834\\
};
\addlegendentry{NLMS (shifted)}

\addplot [color=mycolor3, dashed, line width=1.0pt, mark=x, mark options={solid, mycolor3}]
  table[row sep=crcr]{%
10	-42.8180204271596\\
20	-36.9749038134021\\
45	-29.981832784248\\
90	-23.9545329617284\\
180	-17.9152060830213\\
180	-17.9152060830213\\
360	-11.9184195264112\\
360	-11.9184195264111\\
720	-6.14790796335864\\
};
\addlegendentry{KF}

\addplot [color=mycolor4, line width=1.0pt, mark=x, mark options={solid, mycolor4}]
  table[row sep=crcr]{%
10	-49.218771456297\\
20	-48.2058257675002\\
45	-45.982091495283\\
90	-43.9091635447085\\
180	-37.4292493993351\\
360	-25.7867779402933\\
720	-11.4726594806198\\
};
\addlegendentry{Proposed: $\numIterations=1$}

\addplot [color=mycolor5, line width=1.0pt, mark=x, mark options={solid, mycolor5}]
  table[row sep=crcr]{%
10	-50.2607127953921\\
20	-49.477497108074\\
45	-48.4835463016088\\
90	-45.7835594733879\\
180	-40.5759087349053\\
360	-30.7928379540376\\
720	-16.3770175467772\\
};
\addlegendentry{Proposed: $\numIterations=2$}

\addplot [color=mycolor6, line width=1.0pt, mark=x, mark options={solid, mycolor6}]
  table[row sep=crcr]{%
10	-53.6732873686124\\
20	-51.4881574938392\\
45	-51.1793188896607\\
90	-47.6996008191227\\
180	-43.9204118459246\\
360	-35.3963392648601\\
720	-20.1407475463511\\
};
\addlegendentry{Proposed: $\numIterations=5$}

\addplot [color=mycolor7, line width=1.0pt, mark=x, mark options={solid, mycolor7}]
  table[row sep=crcr]{%
10	-54.8675563683417\\
20	-51.3566590665221\\
45	-52.6998881559716\\
90	-49.728190731694\\
180	-45.8018418860716\\
360	-38.2776191728584\\
720	-23.1140840038385\\
};
\addlegendentry{Proposed: $\numIterations=10$}

\end{axis}
\end{tikzpicture}%
	\caption{Comparisons of single-channel measurements.}
	\label{fig:sysDistOverSpeedsSingleChannel}
\end{figure}

Fig.~\ref{fig:sysDistOverTimeSingleChannel} displays the system distances for a single-channel experiment. As expected from \cite{antweiler2008multi-channel}, the system identification using PSEQ excitation yields lower system distances than using white-noise excitation. In \cite{kuhl2018tracking} it was analyzed that the estimated coefficients, compared to the true coefficients, appear to be systematically shifted by half the period length. To take this systematic shift of $\irLengthEst \numSources / 2$ samples into account, we compute a modified system distance (\begin{tikzpicture} \draw[fill=white, white] (0,0) rectangle (1ex,1ex); \draw[color=mycolor2, dashdotted, line width=1.0pt] (0,0.5ex) -- (1.5em,0.5ex);\end{tikzpicture}) comparing $\vecIr_{\idxTime-\irLengthEst \numSources / 2, \idxSource}$ to $\vecIrEst_{\idxTime, \idxSource}$. This can be considered when mapping HRIRs to specific orientations and effectively reduces the system distance by about \SI{6}{dB}. The KF performance is on par with the NLMS performance. In contrast, the proposed method achieves system distances about \SI{20}{dB} to \SI{30}{dB} lower than the conventional methods in many parts of the signal. At the frame boundaries, every \SI{50}{ms}, discontinuities are observed due to the hard transitions between segments and the parameters learned. Additionally, the time-variance index (TVI) is depicted in Fig.~\ref{fig:sysDistOverSpeedsSingleChannel} as a measure of time-variance between consecutive HRIRs \cite{kuhl2017kalman}, i.e., $\text{TVI} \left( \idxTime \right) = 10 \log_{10} \left( \| \vecIr_{\idxTime, \idxSource} - \vecIr_{\idxTime-1, \idxSource} \|^2 / \|  \vecIr_{\idxTime-1, \idxSource} \|^2  \right)$. Near the so-called bright spot \cite{duda1998range}, the changes between adjacent HRIRs are minor, and the TVI falls steeply below \SI{-60}{dB}. For the conventional methods this lower variation seems to be beneficial, and the proposed method seems inferior here. The parameters learned cannot predict the system change at this specific angle, where the system remains more constant. This might be due to learning the parameters on comparably long segments (\SI{150}{ms}) which contain more variation.

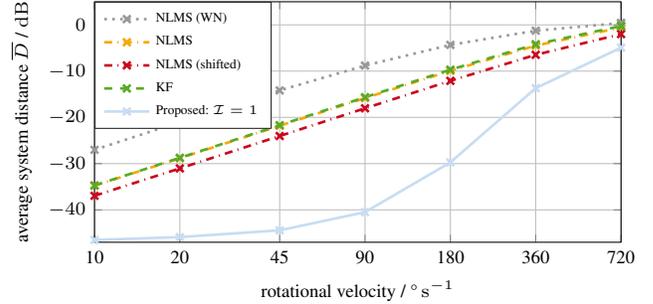
\begin{figure}%
	\centering
	\definecolor{mycolor1}{rgb}{0.96471,0.65882,0.00000}%
\definecolor{mycolor2}{rgb}{0.80000,0.02745,0.11765}%
\definecolor{mycolor3}{rgb}{0.34118,0.67059,0.15294}%
\definecolor{mycolor4}{rgb}{0.78039,0.86667,0.94902}%
\begin{tikzpicture}

\begin{axis}[%
width=0.713\columnwidth,
height=7.5cm,
at={(0\columnwidth,0cm)},
scale only axis,
xmode=log,
xmin=10,
xmax=720,
xtick={10,20,45,90,180,360,720},
xticklabels={{10},{20},{45},{90},{180},{360},{720}},
xminorticks=true,
xlabel style={font=\color{white!15!black}},
xlabel={rotational velocity / $\si{\degree\per\second}$},
ymin=-47,
ymax=5,
ylabel style={font=\color{white!15!black}},
ylabel={average system distance $\sysDistMean$ / \si{dB}},
axis background/.style={fill=white},
xmajorgrids,
xminorgrids,
ymajorgrids,
legend style={at={(0.03,0.97)}, anchor=north west, legend cell align=left, align=left, draw=white!15!black},
footnotesize,legend style={font=\tiny,at={(axis cs: 10, 5)}},width=0.85\columnwidth, line join=round,width=7.0cm,height=3.2cm,,label style={font=\scriptsize},tick label style={font=\scriptsize}
]
\addplot [color=white!60!black, dotted, line width=1.0pt, mark=x, mark options={solid, white!60!black}]
  table[row sep=crcr]{%
10	-27.0244362805797\\
20	-21.0013203408\\
45	-14.1726636963149\\
90	-8.8032262341874\\
180	-4.33336522233774\\
360	-1.26001864599305\\
720	0.388794135874674\\
720	0.388794135874673\\
};
\addlegendentry{NLMS (WN)}

\addplot [color=mycolor1, dashdotted, line width=1.0pt, mark=x, mark options={solid, mycolor1}]
  table[row sep=crcr]{%
10	-34.7652338564382\\
20	-28.819225436588\\
45	-21.8174867846689\\
90	-15.8304243941728\\
180	-9.94005854643916\\
360	-4.48300426360735\\
720	-0.565460853018885\\
};
\addlegendentry{NLMS}

\addplot [color=mycolor2, dashdotted, line width=1.0pt, mark=x, mark options={solid, mycolor2}]
  table[row sep=crcr]{%
10	-36.9563712757428\\
20	-31.029143053932\\
45	-24.031651611527\\
90	-18.037726138521\\
180	-12.1030144014596\\
360	-6.47463099334453\\
720	-1.99874114915676\\
};
\addlegendentry{NLMS (shifted)}

\addplot [color=mycolor3, dashed, line width=1.0pt, mark=x, mark options={solid, mycolor3}]
  table[row sep=crcr]{%
10	-34.7208433491348\\
20	-28.7451574278304\\
45	-21.6883644147081\\
90	-15.6421249382267\\
180	-9.69039420985822\\
360	-4.19009225469724\\
720	-0.243761299448321\\
720	-0.243761299448308\\
};
\addlegendentry{KF}

\addplot [color=mycolor4, line width=1.0pt, mark=x, mark options={solid, mycolor4}]
  table[row sep=crcr]{%
10	-46.4987130366554\\
20	-45.8824367694862\\
45	-44.4247434523605\\
90	-40.4895939489472\\
180	-29.7305067973483\\
360	-13.686874297543\\
720	-4.92194273819931\\
};
\addlegendentry{Proposed: $\numIterations=1$}

\end{axis}
\end{tikzpicture}%
	\caption{Comparisons of three-channel measurements.}
	\vspace{-0.5em}
	\label{fig:sysDistOverSpeeds}
\end{figure}

Fig.~\ref{fig:sysDistOverSpeedsSingleChannel} shows the average system distances for different rotational velocities for $\numSources = 1$. The ranking of the conventional methods is similar to the one in Fig.~\ref{fig:sysDistOverTimeSingleChannel}. As expected, the faster the subject rotation is, the worse the methods perform. The proposed method yields increased performance, which increases with the number of iterations conducted. For $\numIterations = 10$, the average system distance is about \SI{7}{dB} to \SI{22}{dB} lower than in the NLMS-based methods. Further modest improvements are expected for higher numbers of iterations, but a comprehensive analysis exceeds the scope of this contribution. The largest improvements are observed at the rotational velocities \SI[number-unit-product={}]{90}{\degree \per \second} and \SI[number-unit-product={}]{180}{\degree \per \second}. However, the improvements decrease for both lower and higher rotational velocities. We believe that the improvements depend on the segment length $\numObservations$. Better results at the lower and at the higher rotational velocities might be obtained with different segment lengths. The preliminary results in Fig.~\ref{fig:sysDistOverSpeeds} for $\numSources = 3$, with $\numIterations = 1$ iteration to limit the computing time, show similar improvements by up to \SI{20}{dB}. This underlines the potential of the proposed method for multi-channel measurements. All in all, we conclude that learning the system models is advantageous, and we expect this advantage to persist, to some degree, in real-world measurements.

If we assume that $\sysDistMean = \SI{-35}{dB}$ was sufficient for a given application, a maximum rotational velocity of \SI[number-unit-product={}]{10}{\degree \per \second} was possible for the NLMS with white-noise excitation in the single-channel case (cf. Fig.~\ref{fig:sysDistOverSpeedsSingleChannel}). Using PSEQs and considering the systematic shift would allow to increase the velocity to \SI[number-unit-product={}]{45}{\degree \per \second}. However, with $\numIterations = 10$ in the proposed method, the same quality could be obtained at \SI[number-unit-product={}]{360}{\degree \per \second}---a speedup by a factor of 8 to 36. As a result, the measurement duration can be reduced considerably. Notice that the proposed method can also be applied to continuous measurements recorded already.

\section{Conclusion}
\label{sec:conclusion}

In this paper, we have presented a novel method for continuous HRTF measurements. It is based on learning state-space model parameters on short signal segments using the LDS-EM algorithm. Compared to established methods, our method allows to measure HRTFs with considerably lower errors or to significantly reduce the measurement duration. This method paves the way for measuring HRTFs at natural head rotation velocities. Future work will investigate how to optimally choose the segment lengths depending on the rotational velocity. Furthermore, measurements in reverberant rooms and with unconstrained subject movement will be investigated. %

\section{Acknowledgments}
Simulations were performed with computing resources granted by RWTH Aachen University under project rwth0827.

\vfill\pagebreak

\bibliographystyle{IEEEtran}
{\footnotesize
\bibliography{references_selection}
}

\end{document}